# Egret-1: Pretrained Neural Network Potentials for Efficient and Accurate Bioorganic Simulation

**Elias L. Mann**
eli@rowansci.com

**Corin C. Wagen**
corin@rowansci.com

**Jonathon E. Vandezande**

**Arien M. Wagen**

**Spencer C. Schneider**

Rowan Scientific Corporation

## Abstract

Accurate simulation of atomic systems has the potential to revolutionize the design of molecules and materials. Unfortunately, exact solutions of the Schrödinger equation scale as $O(N!)$ and remain inaccessible for systems with more than a handful of atoms, forcing scientists to accept steep tradeoffs between speed and accuracy and limiting the reliability and utility of the resultant simulations. Recent work in machine learning has demonstrated that neural network potentials (NNPs) can learn efficient approximations to quantum mechanics and resolve this tradeoff, but existing NNPs still suffer from limited accuracy relative to state-of-the-art quantum-chemical methods. Here, we present Egret-1, a family of large pretrained NNPs based on the MACE architecture with general applicability to main-group, organic, and biomolecular chemistry. We find that the Egret-1 models equal or exceed the accuracy of routinely employed quantum-chemical methods on a variety of standard tasks, including torsional scans, conformer ranking, and geometry optimization, while offering multiple-order-of-magnitude speedups relative to legacy methods. We also highlight important lacunae for future NNP research to investigate, and suggest strategies for building future high-quality models with increased scale and generality.

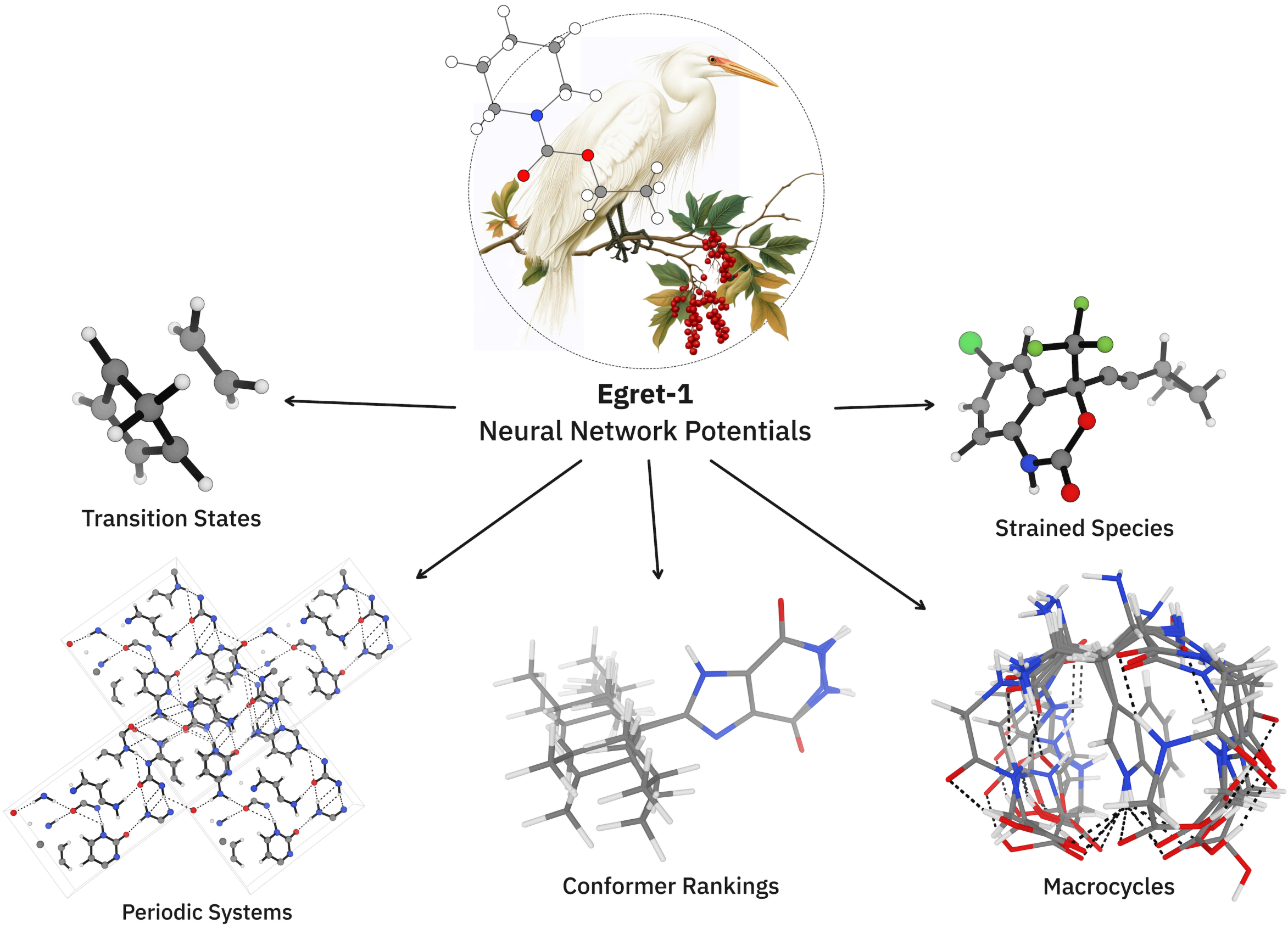

# 1 Introduction

Rational simulation-guided design of atomic systems has been a dream of researchers across the chemical sciences for decades. In principle, fast, accurate, and reliable simulation could lead to vast research accelerations by allowing scientists to replace costly laboratory experiments with fast and inexpensive calculations. High-accuracy prediction of protein–ligand binding affinities would accelerate hit-to-lead optimization in drug discovery and allow limited experimental resources to be allocated more efficiently, while simulation of crystal polymorph landscapes could prevent polymorphism-driven process catastrophes like Abbott's 1998 ritonavir withdrawal.[1–3] In materials science, prediction of material properties could vastly accelerate the search for new polymers, batteries, and carbon-capture materials, and the ability to accurately simulate reaction mechanisms could enable more efficient industrial processes and the design of next-generation catalysts.[2]

Unfortunately, these "holy grails" for computational chemistry and materials science remain largely the domain of science fiction.[2,4] Exact solutions of the Schrödinger equation scale as $O(N!)$ and remain inaccessible for systems with more than a handful of atoms,[5] requiring scientists today to accept steep tradeoffs between speed and accuracy when conducting simulations. While the immense difficulty and importance of this problem has given rise to a plethora of different research programs over the past century, two main approaches have emerged, each with their own advantages and disadvantages.

The first of these approaches, quantum chemistry, seeks to find efficient physically justified approximations to the Schrödinger equation that enable computations to be completed on real systems while maintaining the accuracy and generality of the underlying theory as much as possible. Over the past century, this field has advanced to the point where many molecular properties can now be predicted with greater-than-experimental accuracy and calculations can routinely be run on systems with hundreds of atoms. Yet these approaches are still too slow and costly to model large systems or dynamic processes,[6–8] even though 20–40% of many national supercomputers' time is already spent running quantum chemistry.[9]

The second of these approaches, molecular mechanics, replaces quantum mechanics-based theories altogether with simple classical models fit to reproduce experimental or quantum mechanics-derived values.[10] While these forcefield models allow for many order-of-magnitude speedups relative to quantum chemistry, their limited expressivity makes them unable to describe many complex chemical phenomena. The shortcomings of conventional forcefields have been shown to result in inaccurate predictions of small molecule conformational and torsional preferences, RNA structure, protein-folding dynamics, and hydration free energies.[11–17] Furthermore, most forcefields are by design unable to model reactive processes that involve forming or breaking bonds, limiting their applicability to many important chemical phenomena like catalysis and covalent inhibition.

In recent years, machine-learned models of atomic systems have emerged as a potential resolution to this dilemma. Neural network potentials (NNPs) are machine-learned models trained to reproduce high-level quantum-chemical calculations, typically density-functional theory (DFT), which once trained can mimic the results of quantum chemistry with a single forward-inference step. While early NNPs were trained anew for each specific topology under study, modern NNPs have shown the ability to function as general-purpose forcefields for entire regions of chemical space without system-specific retraining.[18–22] In some cases, NNPs have also been shown to approach the accuracy of the underlying training data while running many orders-of-magnitude faster.[23]

Despite the immense promise of NNPs, their ability to drive vast improvements in atomistic simulation is limited by three main factors:

- **Accuracy and reliability**. Today's NNPs are generally less accurate than conventional quantum chemical simulations, limiting their ability to altogether replace quantum chemical simulations and necessitating specific validation or fine-tuning for each desired application.[24] While what constitutes sufficient accuracy varies considerably from task to task, "chemical accuracy" for simulation is typically defined as an error of less than 1 kcal/mol.[25]
- **Simulation speed**. NNPs remain orders-of-magnitude slower than conventional forcefields despite their considerable speed advantage as compared to quantum chemistry, making it challenging to use NNPs for e.g. molecular-dynamics simulations.[26,27]
- **Generality**. Many NNPs can only simulate a subset of the periodic table and cannot handle charged molecules or unpaired electrons, making it challenging to model many complex processes (although exceptions exist). The best way to handle long-range electrostatic interactions within an NNP framework also remains unclear, although various approaches have been proposed.[21,28–38]

Here, we focus only on the first of these questions, leaving questions of speed and generality to future study. We study the effect of increasing dataset size and diversity for NNPs based on the MACE architecture within the domain of organic and biomolecular chemistry. We also introduce 3 general pretrained models under an MIT license—Egret-1, a general-purpose model for bioorganic simulation; Egret-1e, additionally trained on a variety of main-group structures and superior at thermochemistry; and Egret-1t, additionally trained on transition states—and demonstrate that these models can achieve DFT-level performance for broad regions of chemical space without task-specific fine-tuning, even with conventional datasets and architectures. We highlight the limitations of state-of-the-art NNPs (including the Egret-1 models), and suggest avenues for further improvement.

# 2 Methods

## 2.1 Dataset

In an effort to train more generalizable and accurate models, we compiled a variety of datasets and recomputed all structures at the ωB97M-D3BJ/def2-TZVPPD[39,40] level of theory to match the original MACE-OFF23[41] dataset. In this work, we studied the following datasets, with the caveat that not the entire dataset was added in every case:

- The aforementioned MACE-OFF23 dataset[41] (**M**), which extends the original SPICE[42] dataset of Eastman and coworkers with structures from QMugs[43] and water clusters. All structures are neutral and relatively near-equilibrium.
- Denali (**D**), a subset of the data from the OrbNet Denali[44] dataset which excludes the simple low-lying ChemBL structures, leaving behind (1) protomers and tautomers of ChemBL structures (2) a variety of salt complexes and (3) a set of procedurally generated "exotic" small molecules with unusual bonding patterns.
- Transition1x[45] (**T**), a set of neutral transition states and reactive structures. We employed 100,000 structures from Transition1x, not the entire set.
- Coley3+2[46] (**C**), a set of 3+2 dipolar cycloaddition reaction profiles.
- SPLINTER[47] (**S**), a dataset of dimeric small-molecule interactions mimicking protein–ligand complexes.
- VectorQM24[48] (**V**), an exhaustively enumerated set of ground-state organic and inorganic small molecules.

We also generated an additional dataset ("Finch", **F**) from scratch. We iteratively loaded structures from GDB17[49] & COCONUT[50] and used COATI[51] to generate plausible nearby structures, using agglomerative clustering on the output set to discard similar structures. We generated initial structures for each molecule using the ETKDG[52] algorithm and ran 10 ps of metadynamics using GFN2-xTB[53] with a 1 fs timestep. (The mass of the hydrogen atoms was kept as 1 amu, and the default SHAKE constraints were disabled.) From each output trajectory, 20 dissimilar conformations were selected by agglomerative clustering on heavy-atom RMSD and added to the dataset. (For a subset of the molecules, extra solvent molecules or additives were added to generate complexes of up to four molecules.) We note that metadynamics has previously been proposed as an efficient way to generate training data for NNPs.[54]

For training, we excluded all non-neutral structures and all structures containing a force magnitude greater than 1 Hartree/Å.

We studied many combinations of datasets (*vide infra*), and ultimately selected three for the final Egret-1 models:

- The Egret-1 model is trained on the MACE-OFF23 dataset (**M**). This dataset contains 951,005 total structures and 19,228 unique molecules containing the elements H, C, N, O, F, P, S, Cl, Br, and I.
- The Egret-1e model is trained on the MACE-OFF23 (**M**) and VectorQM24 (**V**) datasets. The combined dataset contains 1,735,879 structures and 98,896 unique molecules containing the elements H, C, N, O, F, Si, P, S, Cl, Br, and I.
- The Egret-1t model is trained on the MACE-OFF23 (**M**), Coley3+2 (**C**), and Transition1x (**T**) datasets. The combined dataset contains 1,079,172 structures and 25,507 unique molecules containing the elements H, C, N, O, F, P, S, Cl, Br, and I.

## 2.2 Model Architecture

The Egret-1 models are derived from the MACE architecture, a high-body-order equivariant message-passing neural network (MPNN) architecture.[55,56] Here we provide a brief explanation of the architecture; for more details, please refer to the original publication.[56]

Graph neural networks are inherently permutation **invariant**, meaning the model's output depends only on the structure of the graph, not on the order in which the atoms and their features are presented. (More generally, a model is invariant to a transformation if applying that transformation to the input of the model does not change the output: $f(G) = f(\mathcal{T}G)$.) Permutation invariance is important for predicting the potential energy of a molecule, since it is dependent on the structural relationships between atoms, not their order.

MACE models are designed to be SO(3) (special orthogonal group in 3 dimensions) **equivariant**, meaning that any rotation in 3D space to the input of the model will rotate the output the same way: $f(QG) = Qf(G)$. (More generally, a model is equivariant to a given transformation if the output of the model transforms predictably when said transformation is applied to the input of a model: $f(\mathcal{T}G) = \mathcal{T}'f(G)$.) SO(3) equivariance is important for accurately predicting directional vectors like atom-centered forces.

The **body order** of an MPNN refers to the number of nodes jointly considered when constructing a message to update the hidden state of a node in a graph. A body order of 2 means that messages are based on pairwise interactions between the central node and each of its neighbors, typically depending only on radial distances. A body order of 3 extends this to include triplet interactions, involving angles formed between the central node and pairs of its neighbors. Increasing the body order allows the MPNN to capture more complex geometric relationships, making the model more expressive, but it also substantially increases the computational cost.

MACE reduces the complexity of computing high-body-order interactions by constructing them from the outer product of all two-body messages. Any regular, local function of the atomic environment can be represented within the span of this outer-product basis, including higher-order relationships between nodes. This approach avoids explicitly computing high-dimensional sums over all body-order combinations. For full proof and derivation, see Ref. (56).

MACE extends the **atomic cluster expansion** (ACE) formulation that any smooth, permutation-invariant and rotation-equivariant function of atomic position can be represented to arbitrary precision as an expansion over body-ordered clusters of atoms.[57] This is used to represent the potential energy of a molecule:

$$E_i = \sum_{Knl} c_{nl}^{(K)} B_{inl}^{(K)} \tag{1}$$

$$E_{\text{total}} = \sum_i E_i \tag{2}$$

where $E_i$ is the energy contribution of atom $i$, $K$ is the body order, $n$ is the radial-basis index and $l$ is the angular degree (spherical-harmonic degree). $B_{inl}^{(K)}$ are the body-ordered basis functions, and $c_{nl}^{(K)}$ are the learned coefficients.

As body order increases and the resolution of the radial and angular bases improves (determining the precision at which distances and angles are represented), the potential energy function can be approximated to arbitrary accuracy. MACE constructs the atomic cluster expansion as

$$E_i = \left( \sum_{t=0}^{T-1} \sum_{\tilde{k}} W_{\text{readout},\,\tilde{k}}^{(t)} h_{i,\tilde{k}00}^{(t)} \right) + \text{MLP}_{\text{readout}}^{(T)} \left( \left\{ h_{ik00}^{(T)} \right\}_k \right) \tag{3}$$

where the term $\sum_{\tilde{k}} W_{\text{readout},\,\tilde{k}}^{(t)} h_{i,\tilde{k}00}^{(t)}$ is analogous to the $c_{nl}^{(K)} B_{inl}^{(K)}$ expansion in ACE: a linear combination of all channels $\tilde{k}$ in the invariant subspace of the hidden state $h$ for atom $i$ at message-passing layer $t$. Here, $l$ and $m$ denote the degree and order of the spherical harmonics, both equal to 0 for invariant terms. Summing the readouts across message-passing layers is analogous to summing over increasing body orders in ACE. An additional multi-layer perceptron (MLP) applied to the final hidden state at layer $T$ serves as a nonlinear correction to capture higher-body-order contributions that are not explicitly computed.

## 2.3 Model Training

Models were trained using PyTorch Lightning,[58] Hydra configuration management,[59] and Distributed Data Parallel for multi-GPU training. We adopted hyperparameters nearly identical to those reported for MACE-OFF23(L), with the exception of removing early stopping and increasing the cutoff radius from 5Å to 6Å.[41] Training was conducted on a cluster of 4 NVIDIA H100 GPUs: Egret-1 required 85 hours to train, Egret-1t required 91 hours, and Egret-1e took 105 hours.

Consistent with the original MACE-OFF23 training strategy, we employed exponential moving average (EMA) of model weight updates throughout training. Each model was trained for 190 epochs, and the loss function combines linearly weighted contributions from both energy and force predictions. For the first 115 epochs, we applied a force loss weight $w_{\text{forces}} = 1000$ and an energy loss weight $w_{\text{energy}} = 40$. During the remaining 65 epochs, stochastic weight averaging (SWA) was used, and the loss weights were adjusted to $w_{\text{forces}}^{\text{SWA}} = 10$ and $w_{\text{energy}}^{\text{SWA}} = 1000$. This training schedule mirrors that of MACE-OFF23 prioritizing force learning during the early phase and shifting focus toward energy accuracy during the averaging phase.[41]

For a full list of hyperparameters, see Table 15 and Table 16 in the Appendix.

To split the dataset into train and validation sets, we grouped structures by molecule to avoid validation-set leakage. Structures were compared by generating a Weisfeiler–Lehman[60] graph hash with NetworkX.[61] Molecules with identical hashes were treated as conformers of the same molecule. To preserve dataset stratification, all structures corresponding to a given molecule were assigned to the same subset. If a molecule appeared in more than one subset, the molecule was assigned to the subset with fewer structures. Subsets were pooled prior to the train-validation split to avoid inter-subset data leakage. For each subset, 80% of the molecules were assigned to the training set and 20% to the validation set. We then computed the ratio for our

train–validation split based on the number of structures to ensure a reliable split; in all cases, the ratio was within 1% of the desired 80/20 train/validation split.

## 2.4 Benchmarking

We evaluate the Egret-1 models against a compendium of theoretical methods commonly used in atomistic simulations, broadly categorized to the unfamiliar reader as follows:

- **Density-functional theory (DFT)**: models electronic structure explicitly via the electron density. Highly accurate but computationally demanding.
- **Semiempirical methods (SE)**: empirically parameterized approximations to quantum mechanics. Much faster than DFT but significantly less accurate.
- **Forcefields (FF)**: classical models using parameterized functions to approximate interatomic forces. Efficient and scalable, but limited in accuracy and transferability.
- **Neural network potentials (NNP)**: machine-learned models trained on quantum-chemical data, offering near-DFT accuracy at forcefield speed.

To more specifically assess the performance of Egret-1 relative to existing NNPs, we benchmark against a set of three high-quality reference models:

- **AIMNet2**, an organic-chemistry-focused NNP trained on a dataset of c. 20M ωB97M-D3BJ/def2-TZVPP molecular calculations that explicitly handles electrostatic and dispersion interactions.[62]
- **MACE-MP-0b2-L**, an NNP using the equivariant MACE architecture trained on calculations of inorganic materials.[63]
- Orb-v3-conservative-inf-omat (abbreviated **Orb-v3** throughout), a new NNP from Orbital Materials which is also trained on calculations of inorganic materials but uses a faster, non-equivariant architecture for improved scaling and inference speed.[64,65]

This list represents, in our view, a balanced set of the best-performing NNPs currently available to researchers, thus highlighting areas in which Egret-1 represents an improvement over the state-of-the-art. We also benchmark against other NNPs for specific benchmarks, like SO3LR[66] and OMat24.[67] Although the MACE-OFF models are the most similar to our work, we are unable to benchmark directly against them owing to license restrictions.[41]

All density-functional-theory computations were conducted with Psi4 1.9.1.[68] For benchmarking, the default settings in Psi4 were modified somewhat: a (99,590) integration grid with "robust" pruning, the Stratmann–Scuseria–Frisch quadrature scheme was employed,[69] and an integral tolerance of $10^{\{-14\}}$ was used throughout. Density fitting was employed for all calculations, and a level shift of 0.10 Hartree was applied to accelerate SCF convergence. For ωB97X-3c calculations, a custom basis-set file was used which adds the missing basis functions for fluorine owing to the documented absence of fluorine in Psi4′s internal implementation of vDZP.[70]

Molecule geometry optimizations were run using geomeTRIC 1.0.2,[71] with the exception of protein optimizations, which were run using FIRE through the Atomic Simulation Environment (ASE).[72] Periodic geometry optimizations were conducted using the ASE QuasiNewton optimizer and a Frechet cell filter, as applicable.

CPU timing studies were conducted on a 12-core Apple M3 Pro with 36 GB RAM, while GPU timing studies were conducted on a single NVIDIA A100 through the Modal cloud platform. Each calculation was run ten times in a row, and the average of the last five runs was taken as the reported time.

# 3 Results

## 3.1 Dataset Sensitivity

We began by testing the effect of different datasets on model performance. For initial model evaluation, we employed a set of benchmarks representative of our desired applications:

- We ran each model through the in-distribution subsets of the GMTKN55[73] molecular thermochemistry benchmark, described in more detail in the "Benchmarks" subsection below. We assess performance through the weighted total mean absolute deviation (WTMAD-2), as is typical for GMTKN55.
- To assess the accuracy of the model's gradient predictions, we compared the gradients predicted by the models to the "ground truth" gradients computed at the ωB97M-D3BJ/def2-TZVPPD level of theory for every molecule in the 1993 Baker[74] optimization set. (The initial non-equilibrium coordinates of every molecule were employed.) Accuracy was assessed via mean cosine similarity.
- Finally, we benchmarked the vibrational frequencies generated by every model on the relevant subset of the VIBFREQ1295[75] dataset, described in more detail in the "Benchmarks" subsection below.

We compared the effect of adding different datasets to the core MACE-OFF23 dataset (Table 1). We observed the most significant improvement in GMTKN55 performance when we added the VectorQM24 dataset (**V**).[48] VectorQM24 is a large set of ground-state main-group structures; adding VectorQM24 to MACE-OFF23 increased the total dataset size from 950K structures to 1.7M structures, and the number of unique structures from 19,228 to 98,896.

| Extra Dataset | Extra Elements | Structures Total | Structures Unique | GMTKN55 | Baker Gradients ↑ | VIBFREQ1295 RMSE ↓ |
|---|---|---|---|---|---|---|
| – | – | 951 001 | 19 228 | 27.17 | 0.9969 | 81.1 |
| **T C** | – | 1 079 172 | 25 507 | 27.12 | 0.9821 | 105.9 |
| **V** | Si | 1 735 879 | 98 896 | 22.34 | 0.9731 | 228.7 |
| **T C V** | Si | 1 864 046 | 100 139 | 23.56 | 0.9589 | 238.0 |
| **V F** | Si | 1 743 430 | 98 996 | 27.67 | 0.9665 | 266.1 |
| **D** | – | 1 241 900 | 32 578 | 63.84 | 0.9814 | 1467.2 |
| **S** | Na | 1 043 464 | 24 462 | 62.30 | 0.8373 | 790.0 |

Table 1: Effect of additional datasets (beyond **M**) on model performance; dataset abbreviations defined more fully above. All datasets contain the elements H, C, N, O, F, P, S, Cl, Br, and I. GMTKN55 results are reported as WTMAD-2, Baker gradient results are reported as mean cosine similarity, and VIBFREQ1295 results are reported in inverse centimeters.

Interestingly, this model performed substantially worse on gradients and frequencies than the MACE-OFF23-only model. We hypothesize that this occurs because all VectorQM24 structures are optimized structures with forces near zero. Since our loss function is a linear combination of force and energy error, we believe that this data may be biasing the model to make force predictions close to zero. This bias degrades gradient predictions, and by extension, frequency predictions. Similar effects have previously been reported by Bowen Deng and co-workers.[76]

We also experimented with adding a variety of transition-state datasets to our baseline dataset. Since the potential-energy surface near a transition state is quite different from other regions of the potential-energy surface, we hypothesized that including transition states in the dataset could dramatically improve model performance. We found that adding structures from Transition1x[45] (**T**) and Coley3+2[46] (**C**) datasets led to improvements on the GMTKN55 transition-state subsets (like BHPERI and BHDIV)—the WTMAD-2 for the GMTKN55 barrier-height subsets decreased from 37.74 to 26.02. However, these improvements were offset by a slightly decreased gradient accuracy and a substantial decrease in the accuracy of frequency predictions. Previous work from Eric Yuan and co-workers has shown that adding transition-state structures to NNP datasets can degrade the quality of Hessian predictions, similar to the effect seen here.[77]

We hypothesized that adding non-equilibrium structures to the VectorQM24 dataset might attenuate the force-related errors observed, and that improved performance on barrier heights might complement the overall superior thermochemistry of the VectorQM24 model. This hypothesis proved to be false. Combining the transition-state datasets with the VectorQM24 dataset led to a model with worse predictions all around than either of the previous models; similarly, adding the highly non-equilibrium Finch dataset to VectorQM24 made the model worse. Efforts to add additional sources of dataset diversity (Denali and SPLINTER) also resulted in decreased model performance, even in domains where the data might presumably be added to help (e.g. non-covalent interactions for SPLINTER).[44,47]

Overall, the results of this dataset study seem to show that, at present margins, increasing dataset diversity leads to decreased overall performance. This conclusion is surprising, and we discuss it further below.

## 3.2 Model Size and Training Length

| **Size** | **Cutoff Radius** | $\ell_{\mathrm{max}}$* | **Channels** | **Parameters** |
|---|---|---|---|---|
| Small | 4.5 | 0 | 96 | 748 000 |
| Medium | 5.0 | 1 | 128 | 1 400 000 |
| Large | 5.0 | 2 | 192 | 3 600 000 |

Table 2: Hyperparameters for each model size for training length study. For final Egret-1 models, the "Large" hyperparameters were used with an increased cutoff radius of 6.0 Å.
*maximum spherical harmonic degree used in the angular basis functions.

| Size | Epochs | GMTKN55 ↓ | Baker Gradients ↑ | VIBFREQ1295 MAE ↓ | RMSE ↓ |
|---|---|---|---|---|---|
| Small | 20 | 27.25 | 0.9930 | 36.1 | 78.7 |
| | 40 | 27.17 | 0.9969 | 35.2 | 81.6 |
| | 190 | 23.58 | 0.9988 | 32.9 | 83.5 |
| Medium | 40 | 24.36 | 0.9992 | 28.9 | 71.2 |
| | 120 | 21.56 | 0.9994 | 28.3 | 75.0 |
| Large | 40 | 23.93 | 0.9997 | 27.3 | 67.7 |
| | 80 | 23.18 | 0.9999 | 27.3 | 69.9 |
| | 120 | 22.17 | 0.9999 | 27.4 | 69.0 |

Table 3: Performance comparison of Egret model sizes across training lengths. GMTKN55 results are reported as WTMAD-2, Baker gradient results are reported as mean cosine similarity, and VIBFREQ1295 results are reported in inverse centimeters.

We next studied the effect of model size and training length. Consistent with the results from MACE-OFF23,[41] we observe that larger models perform better than smaller ones, and training longer improves performance on almost every benchmark. Curiously, we do not see this trend for the VIBFREQ1295 benchmark—in almost every case, the root-mean-squared error increased as training continued.

Based on these results, we selected three models for further study, all employing the "large" architecture: Egret-1, Egret-1e, and Egret-1t. Egret-1 is our recommended general-purpose model, trained on the MACE-OFF23 dataset. Egret-1e, trained on the MACE-OFF23 and VectorQM24 datasets, is best for thermochemistry and can also be used for molecules containing silicon. Finally, Egret-1t is best for transition-state calculations and was trained on the MACE-OFF23, Transition1x, and Coley3+2 datasets.

## 3.3 Thermochemistry and Barrier Heights

We assessed the performance of the Egret-1 models on subsets of the GMTKN55 set, which is commonly employed to benchmark new density functionals.[73] After removing subsets outside of Egret-1′s domain of applicability, 23 of the 55 subsets of GMTKN55 remained. We assessed the WTMAD-2 for various subsets of GMTKN55, as well as the overall WTMAD-2 (Table 4).

| Method | Type | Thermo Chem ↓ | Barrier Heights ↓ | NCI Intra- ↓ | NCI Inter- ↓ | Overall WTMAD-2 ↓ |
|---|---|---|---|---|---|---|
| ωB97M-D3(BJ)[a] | DFT | 5.60 | 2.49 | 3.63 | 4.31 | 4.01 |
| $r^2$SCAN-3c | DFT | 5.56 | 8.64 | 7.11 | 5.67 | 6.67 |
| B97-3c | DFT | 11.05 | 8.67 | 11.93 | 8.69 | 10.39 |
| GFN2-xTB | SE | 21.19 | 17.65 | 11.44 | 24.58 | 19.44 |
| AIMNet2 | NNP | 17.14 | 11.82 | 20.75 | 12.25 | 14.57 |
| MACE-MP-0b2-L | NNP | 20.19 | 15.20 | 20.71 | 36.10 | 27.02 |
| Orb-v3 | NNP | 13.56 | 10.05 | 17.63 | 28.67 | 21.42 |
| OMat24 eqV2-L | NNP | 88.27 | 73.09 | 659.50 | 40.43 | 221.17 |
| SO3LR[b] | NNP | 61.36 | 23.49 | – | 21.66 | – |
| Egret-1 | NNP | 9.45 | 35.61 | 45.45 | 7.72 | 20.91 |
| Egret-1e | NNP | 6.89 | 28.11 | 37.42 | 8.16 | 17.40 |
| Egret-1t | NNP | 9.45 | 13.93 | 45.62 | 10.19 | 19.88 |

Table 4: GMTKN55 results (WTMAD-2) for the 23 in-distribution subsets.
[a]ωB97M-D3(BJ)/def2-QZVP
[b]SO3LR model does not support Br, so the HAL59 subset cannot be run.

For tasks like reaction thermochemistry and intramolecular noncovalent interactions (NCIs), the Egret-1 models far outperform other NNPs and score similarly to quantum-chemical methods like B97-3c. Barrier heights—currently a challenge for NNPs[76,77]—are poorly described by Egret-1 and Egret-1e, which lack transition states in their training data, but adding transition-state data to Egret-1t dramatically improves the model's performance. While the breadth of systems included in GMTKN55 remains a challenge for today's NNPs, these results suggest that increasing the diversity of training data employed provides a path towards increasing NNP generality.

Much of the poor performance of the Egret-1 models on the intramolecular NCI dataset can be ascribed to three challenging datasets: CARBHB12, which examines hydrogen bonding between carbenes and small molecules; PNICO23, which looks at pnictogen bonding; and HAL59, which examines halogen bonding (Figure 2). These exotic interactions are poorly described in the training data, leading to poor model performance.

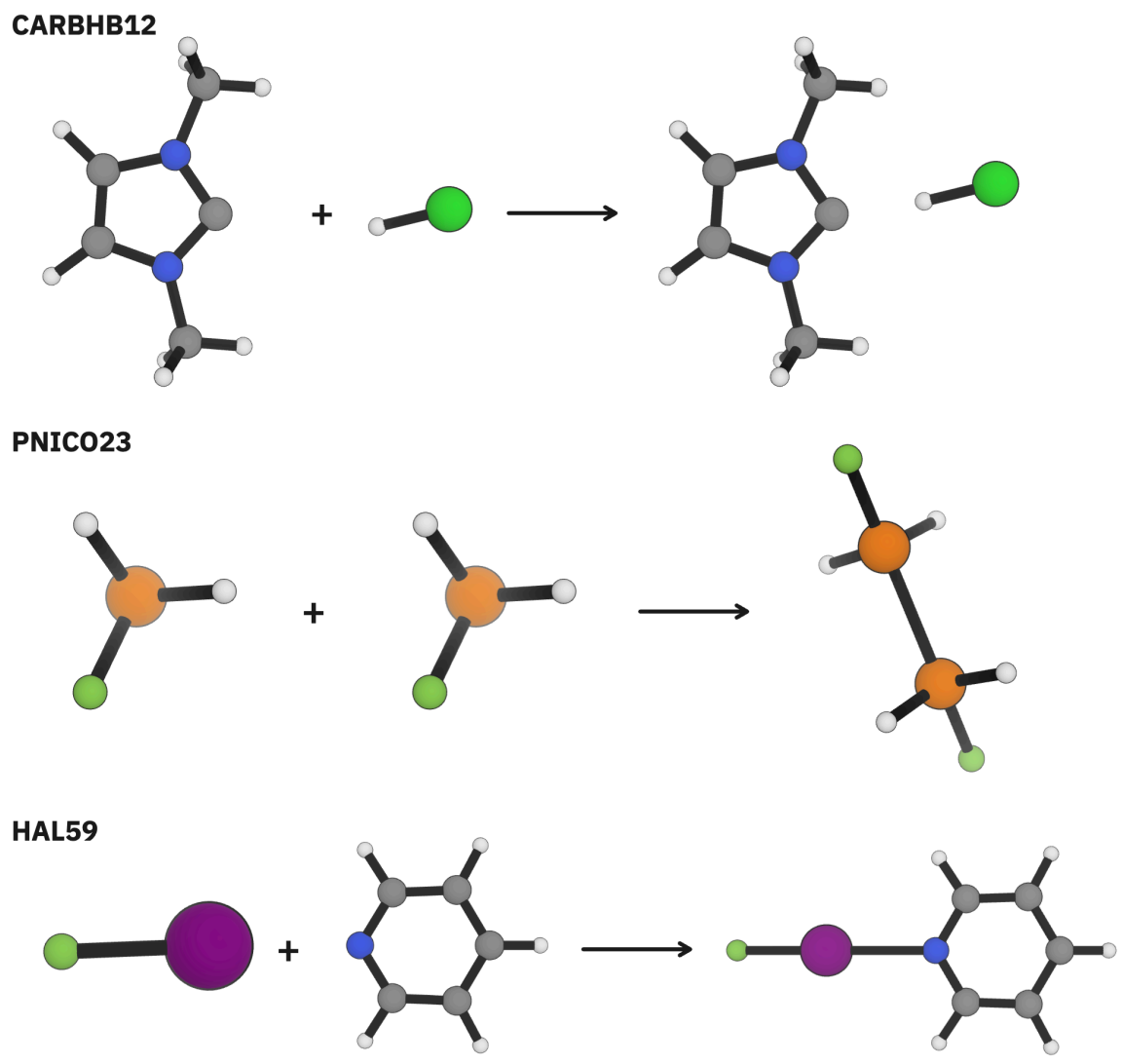

Figure 2: Several particularly challenging structures from the intermolecular non-covalent interaction subsets of GMTKN55.

### 3.4 Molecular Geometries

To see if the Egret-1 models could be used to generate accurate molecular geometries, we benchmarked them against the ROT34 benchmark set.[78] ROT34 tests the ability of computational methods to generate accurate gas-phase geometries through geometry optimization, as quantified by agreement with gas-phase rotational constants. Rotational constants are an exquisite probe of molecular geometry, as accurate prediction of rotational constants indicates that bond lengths, angles, and dihedral constants can all be reproduced correctly. The three Egret-1 models performed very well on ROT34, with mean absolute deviation and maximum deviation smaller than any other method surveyed, including composite density-functional-theory methods like r²SCAN-3c (Table 5).

| **Method** | **Theory** | **MD (%)** | **MAD (%) ↓** | **MAX (%) ↓** | **SD (%) ↓** |
|---|---|---|---|---|---|
| ωB97X-3c | DFT | −0.32 | 0.38 | 1.05 | 0.34 |
| r²SCAN-3c | DFT | 0.71 | 0.75 | 1.36 | 0.41 |
| B97-3c | DFT | 0.36 | 0.51 | 1.44 | 0.54 |
| GFN2-xTB | SE | −1.54 | 2.85 | 24.82 | 6.59 |
| OpenFF Sage | FF | 3.00 | 3.13 | 9.32 | 2.07 |
| AIMNet2 | NNP | 0.08 | 0.48 | 2.71 | 0.67 |
| MACE-MP-0b2-L | NNP | 2.01 | 2.11 | 5.67 | 1.19 |
| Orb-v3 | NNP | 1.36 | 1.41 | 6.58 | 1.19 |
| Egret-1 | NNP | 0.15 | 0.21 | 0.59 | 0.22 |
| Egret-1e | NNP | 0.16 | 0.24 | 0.52 | 0.22 |
| Egret-1t | NNP | 0.10 | 0.21 | 0.69 | 0.25 |

Table 5: ROT34 rotational constants benchmark results, reported as mean deviation (MD), mean absolute deviation (MAD), maximum deviation (MAX), and standard deviation (SD).

## 3.5 Vibrational Frequencies

We next sought to assess the accuracy of the Hessian matrices predicted by the Egret-1 models. Towards this end, we benchmarked the models against the VIBFREQ1295 dataset.[75] After excluding molecules outside of Egret-1's domain of applicability, 115 molecules remained with reference vibrational frequencies computed at the CCSD(T)(F12*)/cc-pVDZ-F12 level of theory. We tested a variety of different levels of theory against this benchmark (Table 6).

| **Method** | **Theory** | **MAE ↓** | **RMSE ↓** |
|---|---|---|---|
| ωB97X-3c | DFT | 41.0 | 53.5 |
| $r^2$SCAN-3c | DFT | 22.3 | 31.1 |
| B97-3c | DFT | 32.6 | 44.5 |
| GFN2-xTB | SE | 66.4 | 86.7 |
| AIMNet2 | NNP | 31.8 | 45.8 |
| MACE-MP-0b2-L | NNP | 94.2 | 110.6 |
| Orb-v3 | NNP | 64.4 | 76.7 |
| Egret-1 | NNP | 24.4 | 49.2 |
| Egret-1e | NNP | 72.3 | 161.9 |
| Egret-1t | NNP | 39.3 | 91.5 |

Table 6: VIBFREQ1295 vibrational frequency benchmark results ($cm^{-1}$).

The three different Egret-1 models performed very differently on this benchmark: while Egret-1e performed poorly (similar to other low-cost methods), Egret-1t performed approximately as well as many commonly employed density functionals, and Egret-1 had one of the lowest mean errors of any method studied, exceeded only by the "Swiss-army-knife" composite DFT method $r^2$SCAN-3c.

As discussed previously (*vide supra*), these results highlight the sensitivity of Hessian predictions to the precise dataset employed for training. Previous work from Eric Yuan and co-workers[77] demonstrating that fine-tuning NNPs on transition-state geometries can result in a consistent underestimation of Hessian eigenvalues, which may explain the gap in performance between Egret-1 and Egret-1t. Similarly, Egret-1e adds a large number of equilibrium structures to the dataset, which may systematically bias the Hessian and gradient predictions. Alternative training strategies, like explicitly training to Hessian data,[79] may ameliorate this sensitivity in the future.

## 3.6 Torsional Profiles

Accurate prediction of torsional profiles is an important and well-studied task in computer-assisted drug design.[80–82] We evaluated a variety of methods on the TorsionNet206 dataset, which comprises high-level CCSD(T)/def2-TZVP torsional scans for a library of drug-like fragments. We found that the Egret-1 models performed very well relative to other low-cost methods, and even exceeded the performance of common DFT methods like B3LYP-D3BJ/6-31G(d) and $r^2$SCAN-3c. Given the clear effect of accurate torsional parameterization on the accuracy of free-energy-perturbation workflows[80], we expect that Egret-1 can serve as a low-cost oracle method for torsional parameterization workflows like BespokeFit[83] moving forward.

| Method | Theory | MAE ↓ | RMSE ↓ | $R^2$ ↑ | Spearman ↑ |
|---|---|---|---|---|---|
| ωB97M-D3BJ/def2-TZVPPD | DFT | 0.15 | 0.18 | 0.99 | 0.98 |
| B97-3c | DFT | 0.35 | 0.45 | 0.98 | 0.97 |
| $r^2$SCAN-3c | DFT | 0.42 | 0.54 | 0.97 | 0.97 |
| B3LYP-D3BJ/6-31G(d) | DFT | 0.57 | 0.71 | 0.95 | 0.94 |
| GFN2-xTB | SE | 0.73 | 0.91 | 0.85 | 0.85 |
| AIMNet2 | NNP | 0.39 | 0.48 | 0.95 | 0.94 |
| MACE-MP-0b2-L | NNP | 1.15 | 1.43 | 0.74 | 0.75 |
| Orb-v3 | NNP | 0.97 | 1.20 | 0.83 | 0.83 |
| OMat24 eqV2-L | NNP | 1.48 | 1.84 | 0.77 | 0.81 |
| Egret-1 | NNP | 0.20 | 0.24 | 0.99 | 0.98 |
| Egret-1e | NNP | 0.22 | 0.28 | 0.99 | 0.98 |
| Egret-1t | NNP | 0.23 | 0.29 | 0.99 | 0.98 |

Table 7: TorsionNet206 benchmark results; MAE and RMSE reported in kcal/mol, alongside $R^2$ and Spearman correlation coefficients.

## 3.7 Conformers

Flexible molecules exist in a variety of different conformations. Predicting the energy differences between conformers is exceedingly difficult, as it is dominated by subtle changes in strain, solvation, and van der Waals forces. Accurately determining the ground-state conformer and the Boltzmann distribution of conformers is vital for predicting strain energy in docked poses, conformationally averaged properties, and downstream reaction modeling.[84–87]

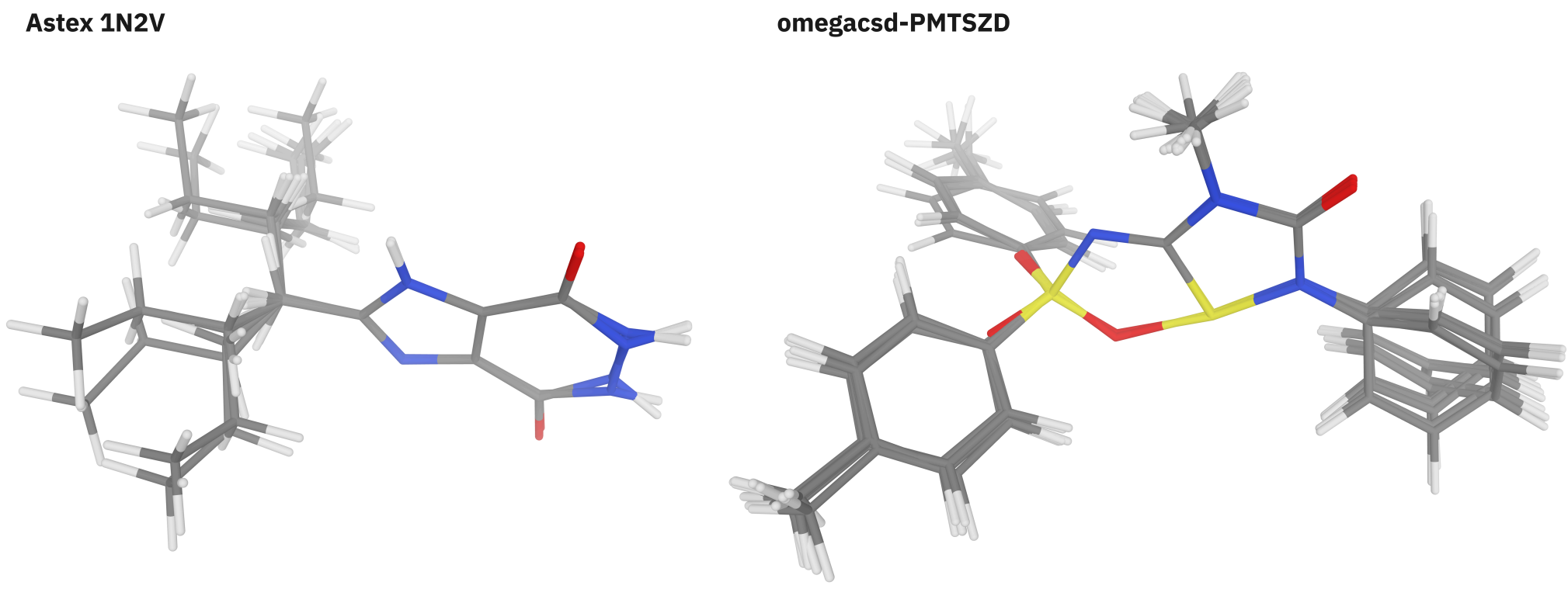

Figure 3: Representative conformer ensembles from the Folmsbee benchmark set.

We employed three different conformer-based benchmarks to assay the performance of the Egret-1 models at this important task. We first benchmarked Egret-1 on the large Folmsbee conformer set,[12] which comprises 708 distinct conformer ensembles computed at the DLPNO-CCSD(T) level of theory (Figure 3). After removing charged species and ensembles for which the DLPNO calculations were not completed, 593 of the 708 ensembles remained. We found that the Egret-1 models had similar performance to DFT methods like B97-3c and far outperformed existing NNPs like AIMNet2, Orb-v3, and MACE-MP-0b2-L (Table 8).

| Method | Theory | Overall MAE↓ | Overall RMSE↓ | Mean $R^2$↑ | Mean Spearman↑ | Incomplete Subsets |
|---|---|---|---|---|---|---|
| ωB97X-D/def2-TZVP | DFT | 0.24 | 0.37 | 0.84 | 0.85 | 4 |
| B3LYP/def2-TZVP | DFT | 0.25 | 0.39 | 0.84 | 0.85 | – |
| B97-3c | DFT | 0.30 | 0.49 | 0.81 | 0.82 | 1 |
| GFN2-xTB | SE | 0.71 | 1.29 | 0.57 | 0.60 | – |
| AIMNet2 | NNP | 0.54 | 0.93 | 0.64 | 0.65 | – |
| MACE-MP-0b2-L | NNP | 1.08 | 2.06 | 0.46 | 0.37 | – |
| Orb-v3 | NNP | 0.88 | 1.71 | 0.51 | 0.50 | – |
| OMat24 eqV2-L | NNP | 0.87 | 1.44 | 0.47 | 0.50 | – |
| Egret-1 | NNP | 0.31 | 0.51 | 0.78 | 0.81 | – |
| Egret-1e | NNP | 0.30 | 0.50 | 0.79 | 0.81 | – |
| Egret-1t | NNP | 0.34 | 0.59 | 0.76 | 0.79 | – |

Table 8: Folmsbee conformer-energy benchmark; MAE and RMSE reported in kcal/mol.

The Folmsbee conformer set focuses on low-energy conformers: the vast majority of Folmsbee conformers are within 5 kcal/mol of the ground state. A contrasting conformer benchmark is Wiggle150,[23] which comprises 150 highly strained conformers of three organic molecules (average relative energy of 103 kcal/mol) and tests the ability of NNPs to handle unusual bond distances and angles (Figure 4).

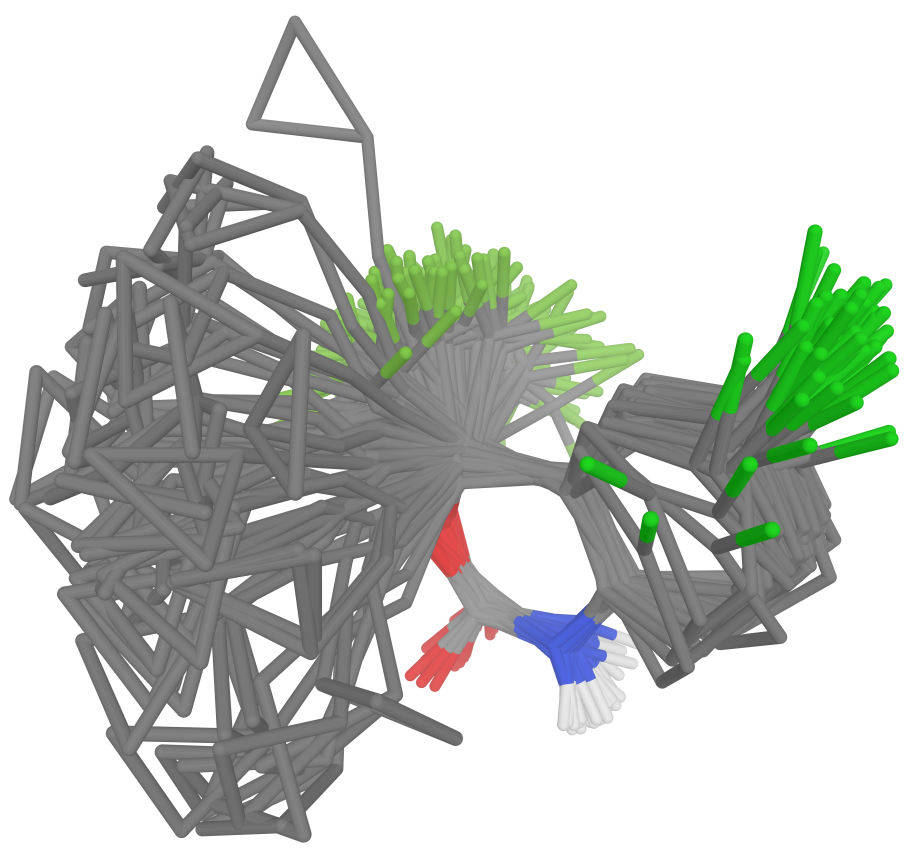

Figure 4: Efavirenz conformers from the Wiggle150 benchmark set.

We evaluated the performance of Egret-1 and other NNPs against Wiggle150 (Table 9). All Egret-1 models performed better than any other NNP surveyed on Wiggle150, and even outperformed most DFT methods: only double hybrids or range-separated hybrids with quadruple-ζ basis sets scored better than the best Egret-1 models.

| Method | Theory | MAE ↓ | RMSE ↓ |
|---|---|---|---|
| ωB97M-D3BJ/def2-QZVP | DFT | 1.18 | 1.59 |
| ωB97x-3c | DFT | 4.12 | 4.63 |
| $r^2$SCAN-3c | DFT | 1.72 | 2.19 |
| B97-3c | DFT | 2.32 | 2.96 |
| B3LYP-D3BJ/6-31G(d) | DFT | 3.46 | 4.01 |
| GFN2-xTB | SE | 14.60 | 15.20 |
| Sage 2.2.1 | FF | 27.20 | 34.60 |
| AIMNet2 | NNP | 2.35 | 3.11 |
| MACE-MP-0b2-L | NNP | 14.59 | 16.37 |
| Orb-v3 | NNP | 7.72 | 8.82 |
| OMat24 eqV2-L | NNP | 6.36 | 7.73 |
| SO3LR | NNP | 10.36 | 12.36 |
| Egret-1 | NNP | 1.58 | 2.15 |
| Egret-1e | NNP | 1.56 | 2.07 |
| Egret-1t | NNP | 1.71 | 2.25 |

Table 9: Wiggle150 strained-conformer benchmark; MAE and RMSE reported in kcal/mol.

Finally, we tested Egret-1's ability to handle complex bioorganic structures through Řezáč's MPCONF196 benchmark, which comprises 196 conformers of 13 complex macrocyclic therapeutics (Figure 5).[88] Organic macrocycles can exhibit exceedingly complex conformational behavior[89–92] and present a particularly challenging test for computational methods—but accurate description of conformational behavior is critical to rational design of macrocyclic therapeutics.[93,94] Egret-1 far outperformed all other NNPs on this benchmark, with an MAE and RMSE comparable to density-functional-theory methods with quadruple-$\zeta$ basis sets (Table 10).

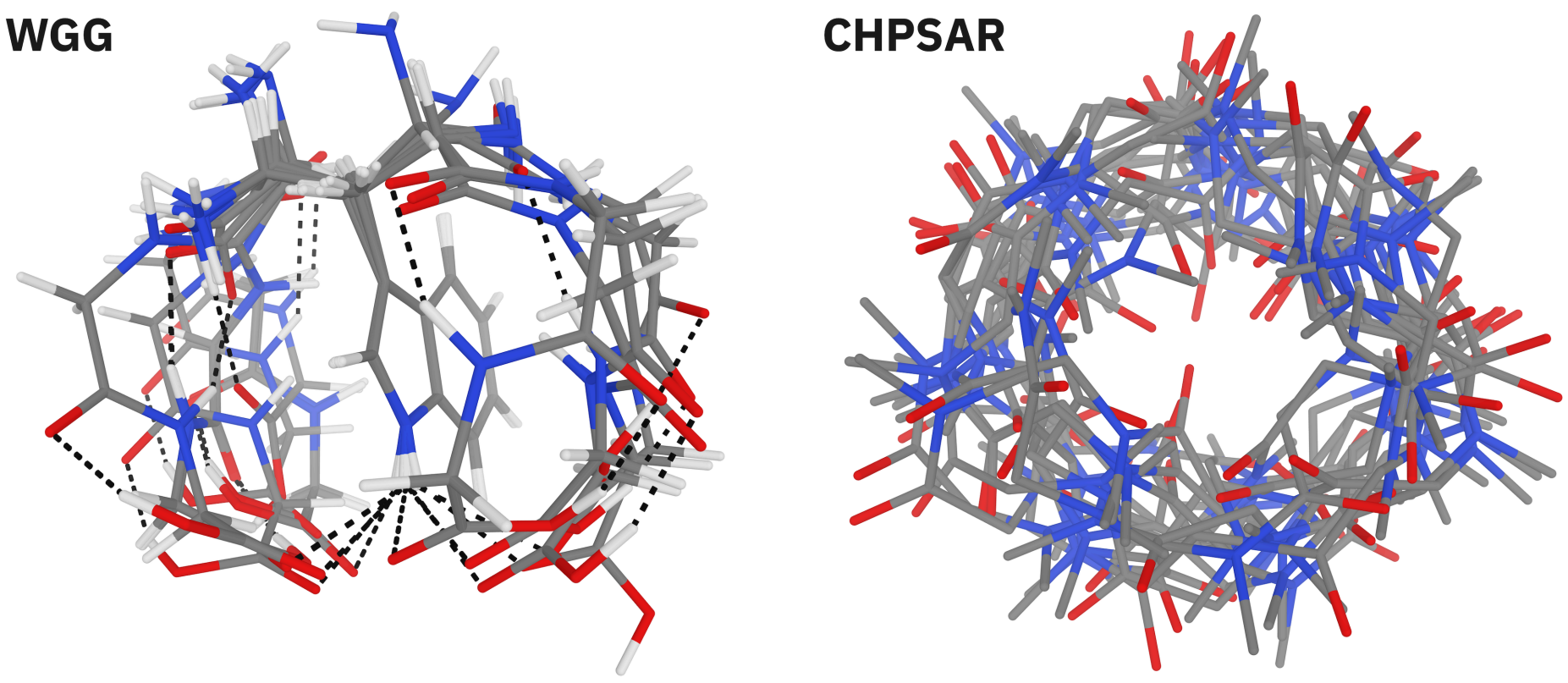

Figure 5: Representative macrocyclic conformer ensembles from the MPCONF196 benchmark set.

| Method | Theory | MAE ↓ | RMSE ↓ |
|---|---|---|---|
| ωB97X-D3/def2-QZVP | DFT | 0.64 | 1.06 |
| B3LYP-D3BJ/def2-QZVP | DFT | 0.62 | 0.92 |
| M06-2X/def2-QZVP | DFT | 1.56 | 2.23 |
| GFN2-xTB | SE | 2.21 | 3.14 |
| AIMNet2 | NNP | 2.06 | 2.71 |
| MACE-MP-0b2-L | NNP | 4.91 | 6.42 |
| Orb-v3 | NNP | 3.94 | 5.32 |
| Egret-1 | NNP | 0.70 | 1.12 |
| Egret-1e | NNP | 1.03 | 1.50 |
| Egret-1t | NNP | 0.93 | 1.45 |

Table 10: MPConf196 macrocycle conformer-energy benchmark; MAE and RMSE reported in kcal/mol.

## 3.8 Molecular Dynamics

Accurate prediction of many chemical phenomena requires molecular dynamics (MD). Unfortunately, good performance on energy- and force-based benchmarks is not enough to guarantee that an NNP will be able to produce stable, well-defined MD trajectories; in 2022, Xiang Fu and co-workers showed that "many existing models are inadequate when evaluated on simulation-based benchmarks, even when they show accurate force prediction" and argued that direct MD-based testing should be employed in the future.[95] In the subsequent years, short MD simulations have become a quick and practical way to assay if a given NNP leads to stable simulations or causes energy leaks and runaway heating.[96–98]

We evaluated the stability of the Egret-1 models on MD simulations of maraviroc, an anti-retroviral therapeutic emblematic of the complex drug-like molecules Egret-1 is intended to study. Simulations were run in the NVT ensemble for 1 ps using a second-order Langevin thermostat,[99] and then propagated for 100 ps in the NVE ensemble using velocity Verlet integration with a 1 fs timestep. We found that all Egret-1 models were stable for the entire 100 ps studied. While further work is needed to assess MD stability on a more varied set of tasks, this experiment rules out the catastrophic energy-leakage scenarios characteristic of previous generations of NNPs.

## 3.9 Periodic Systems

We next investigated if the Egret-1 models were capable of extrapolation to periodic systems. Today, the divide between molecular and periodic quantum chemistry has given rise to two largely separate ecosystems of researchers, software, and publications.[100] Hybrid density functionals, although typically more accurate[101,102] than non-hybrid density functionals for molecular systems, are impractical to employ for all but the smallest periodic systems;[103] most periodic calculations are instead run with non-hybrid density functionals, which is known to create substantial errors in *inter alia* small-molecule conformations.[104]

Graph-based NNPs provide a potential way to connect these two paradigms. Previous work from Kästner[105] and Daru[106] suggests that NNPs trained on high-accuracy molecular DFT calculations might be able to extrapolate to periodic systems, thus providing scientists with a way to extend the accuracy of hybrid molecular DFT to periodic systems. To assess the accuracy of the Egret-1 models at simulating periodic systems, we employed the X23b benchmark set,[107,108] which tests the ability of computational methods to reproduce experimental lattice energies and unit-cell volumes for 23 organic molecular crystals (Table 11).

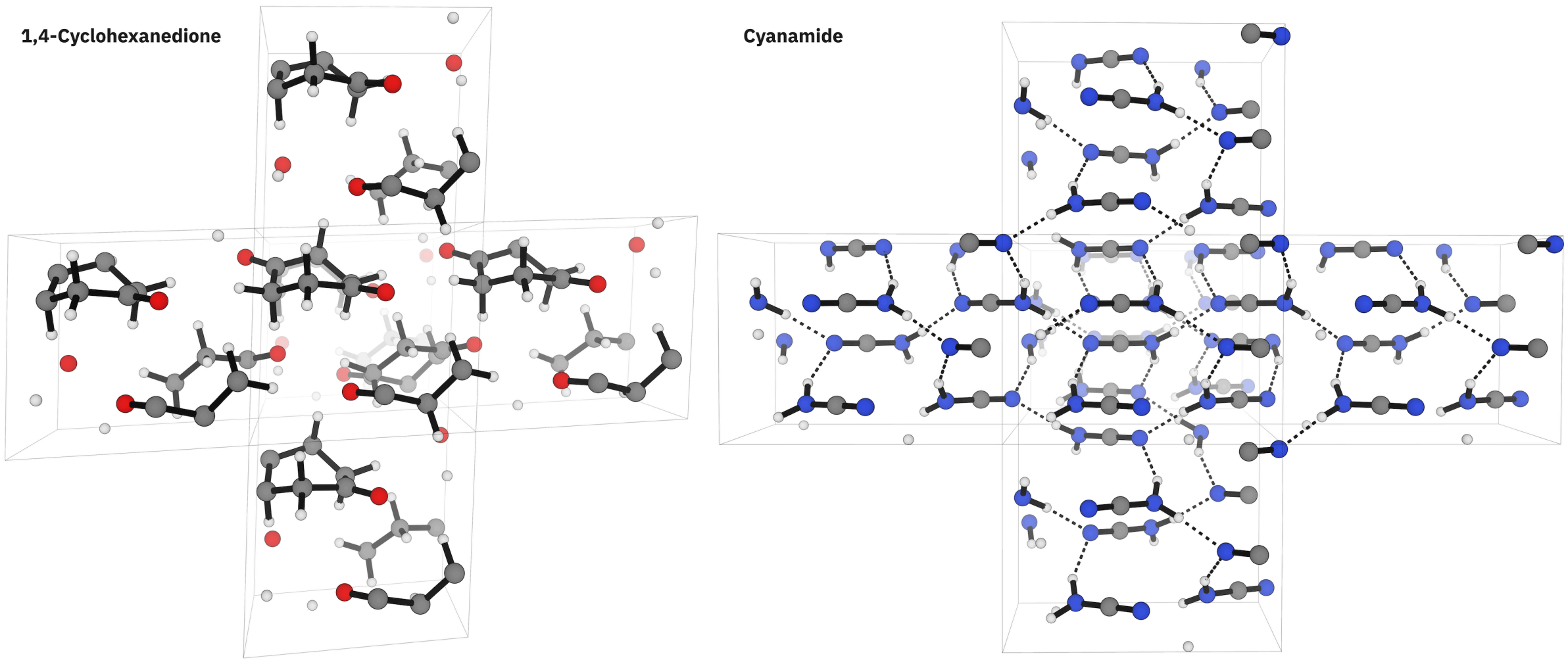

Figure 6: 1,4-cyclohexanedione and cyanamide structures from the X23b benchmark set.

While Egret-1e and Egret-1t performed very poorly on this benchmark set, we found that Egret-1 was the best-performing NNP studied, with a mean-absolute-error for lattice energy of 2.61 kcal/mol. Nevertheless, a considerable gap between low-cost methods and density-functional theory still exists for this task, indicating that achieving chemical accuracy for periodic calculations remain an unsolved challenge for these NNPs.

| Method | Theory | Lattice Energy MAE ↓ | Cell Volume MAPE (%) ↓ |
|---|---|---|---|
| $r^2$SCAN-3c | DFT | 0.97 | 3.68 |
| PBE-D4/QZ//PBE-D4/mTZ | DFT | 0.87 | 4.44 |
| GFN2-xTB | SE | 5.38 | 7.76 |
| AIMNet2 | NNP | 9.44 | 18.52 |
| MACE-MP-0b2-L (with D3BJ) | NNP | 3.47 | 3.03 |
| MACE-MP-0b2-L (no D3BJ) | NNP | 9.28 | 22.39 |
| Orb-v3 | NNP | 26.79 | 12.30 |
| Egret-1 | NNP | 2.61 | 4.38 |
| Egret-1e | NNP | 18.66 | 14.73 |
| Egret-1t | NNP | 30.19 | 12.11 |

Table 11: X23b molecular-crystal benchmark. Lattice energy is reported as mean absolute error (MAE) in kcal/mol; cell volume is reported as mean absolute percent error (MAPE).

Prediction of relative crystal-polymorph stability is crucial in drug formulation; in pathological cases like Abbott's anti-HIV drug ritonavir, the presence of unexpected low-energy polymorphs can lead to catastrophic manufacturing failure and hundreds of millions of dollars in estimated losses.[1,109] An Abbott postmortem concluded that it was "highly advisable... to carry on exhaustive research to identify the most stable and all possible polymorphs" to prevent future ritonavir-level crises,[1] and consequently experimental and computational methods for crystal-polymorph exploration have become key technologies in small-molecule process research and development, even though the requisite periodic DFT calculations typically require millions of CPU-hours on high-performance computing clusters.[110]

To assess Egret-1's ability to accurately identify low-energy crystal polymorphs, we evaluated it on a recent set of organic crystal polymorphs collected by Schrödinger and ranked with the r²SCAN functional.[111] After excluding the pathological PULWIF structure, which led to erratic results with all methods surveyed, 63 sets of crystal polymorphs remained in our benchmark set (Table 12). Energy- and ranking-based benchmarks were computed separately for all 63 sets and then averaged to give the final values shown below.

| **Method** | **MAE ↓** | **RMSE ↓** | **$R^2$ ↑** | **Spearman ↑** |
|---|---|---|---|---|
| AIMNet2 | 2.07 | 2.43 | 0.27 | 0.30 |
| MACE-MP-0b2-L | 0.74 | 0.91 | 0.27 | 0.28 |
| Orb-v3 | 0.76 | 0.94 | 0.25 | 0.26 |
| Egret-1 | 0.76 | 0.97 | 0.34 | 0.33 |

Table 12: Schrödinger crystal-polymorph-ranking benchmark; MAE and RMSE reported in kcal/mol, alongside $R^2$ and Spearman correlation coefficients.

While AIMNet2 performs poorly on this benchmark, the other three NNPs all achieve good relative accuracy here (as assessed by MAE and RMSE), but the large number of accessible low-lying polymorphs leads to poor Pearson and Spearman correlation values. We note that as described above,[104,112] non-hybrid functionals often struggle with relative conformer energies, so the reference r²SCAN data may be somewhat inaccurate—the uniform c. 0.7 kcal/mol MAE may reflect a limitation of this benchmark set. Future work can study if methods like Egret-1 can be integrated into end-to-end crystal-structure-prediction workflows to increase efficiency and accuracy; recent results from Schrödinger indicate that NNPs can be gainfully employed as an intermediate filtering step before running full DFT calculations.[111]

## 3.10 Timing

A key advantage of NNPs relative to conventional quantum-chemical methods like DFT is the potential for dramatic speed increases. To assess the magnitude of the speedups possible with the Egret-1 models, we measured the speed of Egret-1 as compared to other levels of theory (Table 13). While Egret-1 is relatively slow compared to NNPs without equivariance or with fewer parameters, it remains significantly faster than even low-accuracy DFT methods with double-$\zeta$ basis sets like B3LYP/6-31G(d).

| Method | Theory | Compute | Ibuprofen | Citalopram | Rapamycin | Insulin |
|---|---|---|---|---|---|---|
| B3LYP/6-31G(d) | DFT | CPU | 17.34 | 176.88 | 867.14 | Failed |
| GFN2-xTB | SE | CPU | 0.05 | 0.08 | 0.42 | Failed |
| AIMNet2 | NNP | CPU | 0.05 | 0.14 | 0.07 | 0.22 |
| | | GPU | 0.08 | 0.08 | 0.08 | 0.08 |
| MACE-MP-0b2-L | NNP | CPU | 0.30 | 0.67 | 0.96 | 5.62 |
| | | GPU | 0.28 | 0.29 | 0.28 | 0.43 |
| Orb-v3 | NNP | CPU | 0.51 | 0.54 | 0.67 | 5.42 |
| | | GPU | 0.27 | 0.28 | 0.31 | 0.28 |
| Egret-1t | NNP | CPU | 0.81 | 1.31 | 1.80 | 12.50 |
| | | GPU | 0.85 | 0.86 | 1.01 | 1.09 |

Table 13: Time to compute a single-point energy by level of theory, in seconds.

Egret-1 also does not suffer from the SCF- and band-gap-related pathologies of DFT for large biomolecular systems,[113] allowing for smooth and well-behaved optimizations to be easily conducted even on small proteins. Following sanitization and removal of water molecules, we were able to optimize an all-atom structure of human insulin (PDB: 3I40)[114] using the FIRE optimizer in 140 seconds and 716 optimization steps on a single NVIDIA H100 (Figure 7).[115]

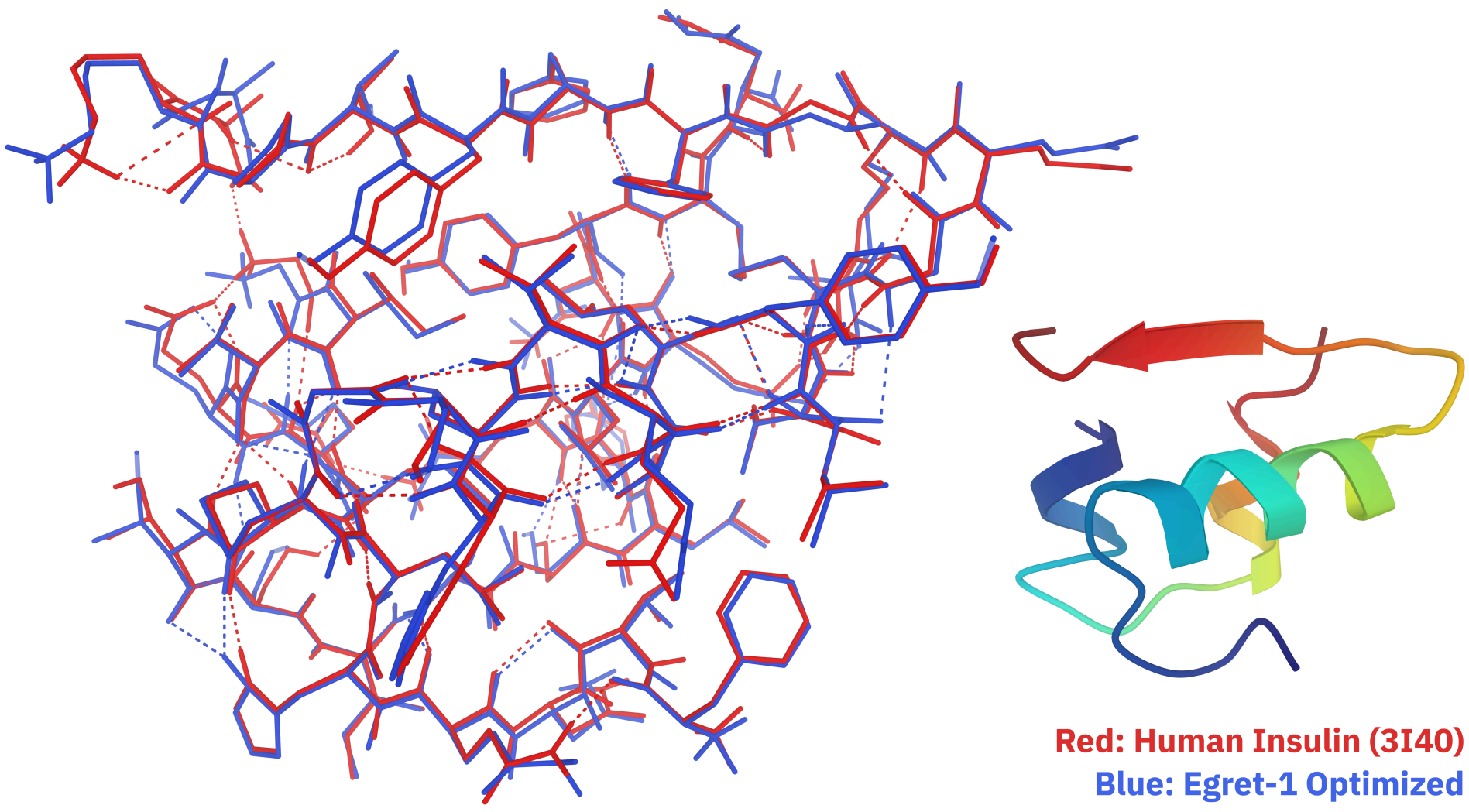

Figure 7: Comparison of initial (red) and optimized (blue) all-atom human insulin structures.

## 3.11 Catalysis

NNPs offer the promise of modeling larger systems than can currently be accessed with DFT with significantly fewer computational resources. In 2017, Daniel DiRocco and co-workers at Merck reported the development of a multifunctional organocatalyst for stereoselective prodrug assembly. Kinetic studies demonstrated catalyst cooperativity, leading DiRocco and co-workers to develop a preorganized dimeric catalyst that displayed significantly increased activity. As a part of this study, the authors found transition states with DFT for these large supramolecular assemblies, reporting a $\Delta\Delta G^\ddagger$ for the two isomers of 2.3 kcal/mol for their initial catalyst and 2.6 kcal/mol for the linked dimeric catalyst. While the size of this system meant that even a single-point calculation with the ωB97X-3c DFT functional took over two hours, reoptimization of the transition state with Egret-1 could be run in under a minute on a consumer-grade NVIDIA RTX 4090 GPU. Egret-1 showed a highly conserved structure during re-optimization, but like all of the tested low-cost methods, overstabilized the lower-energy TS (Table 14).

| Method | Theory | ΔΔG‡ (Separate) | ΔΔG‡ (Linked) |
|---|---|---|---|
| Reference[a] | DFT | 2.29 | 2.6 |
| GFN2-xTB | SE | 3.09 | 4.72 |
| AIMNet2 | NNP | 3.92 | 10.17 |
| Egret-1 | NNP | 5.72 | 6.37 |
| Egret-1e | NNP | 8.87 | 10.33 |
| Egret-1t | NNP | 7.96 | 11.25 |

Table 14: Gibbs-free-energy difference between diastereomers of the transition state. [a]M06L/6-31+G(d,p)//B3LYP-D3/6-31G(d,p)

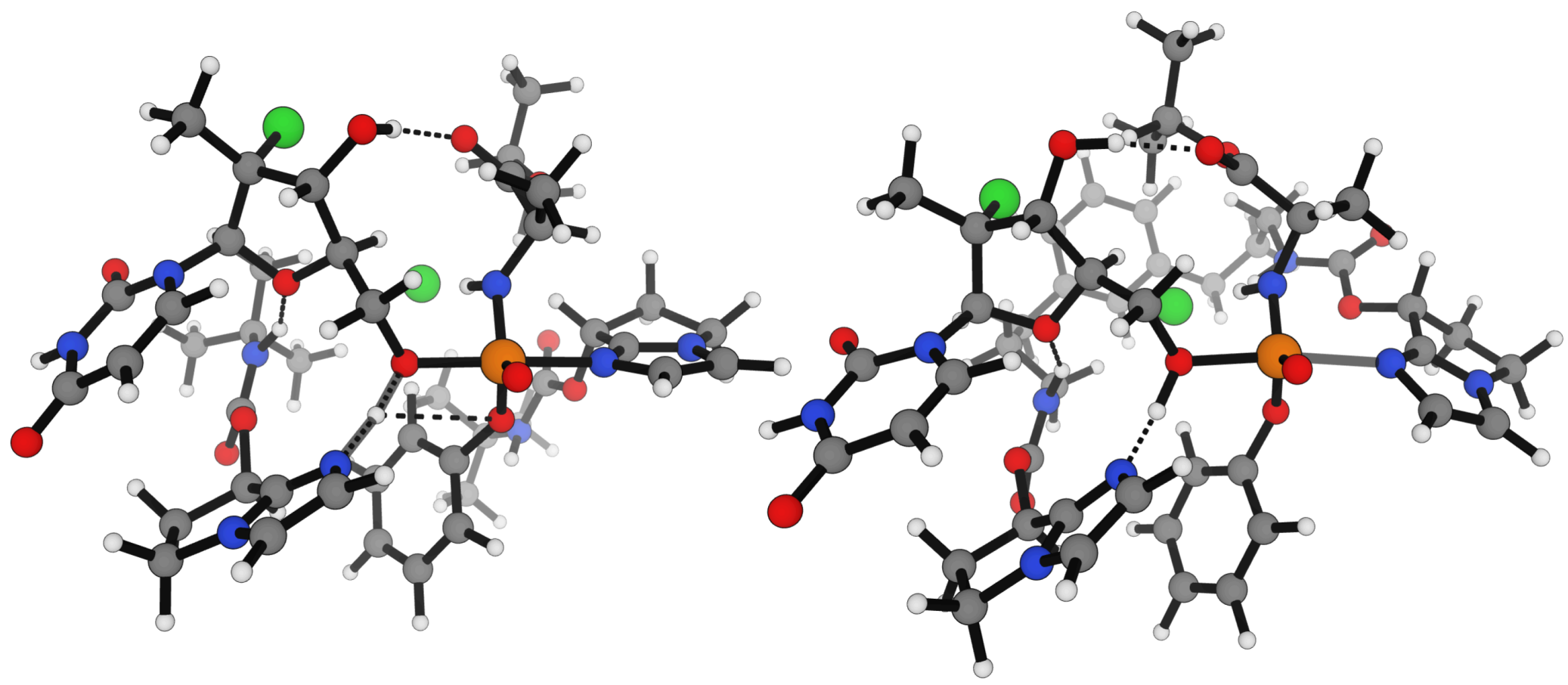

Figure 8: Egret-1 computed structures of the DiRocco transition states for the separate catalysts (left) and linked dimeric catalyst (right).

# 4 Discussion

In this work, we have shown that the Egret-1 models are capable of serving as drop-in replacements for DFT in a variety of applications important to drug discovery and materials science. Surprisingly, no massive fundamental advance in dataset scale or model architecture was necessary to achieve this result—we simply combine previously reported datasets with a well-studied model architecture, with a focus on maximizing accuracy.

Our models occupy a distinct niche in the emerging landscape of ML-based atomistic simulation methods. We focus on maximizing zero-shot accuracy versus experimental or high-level quantum-chemical benchmarks, not on inference speed or data efficiency; accordingly, we anticipate that the near-term use of the Egret-1 models will center around using them as fast surrogates for DFT, not as replacements for forcefields. This focus differs from other recent work in the field, like the optimized Orb models from Mark Neumann and coworkers at Orbital Materials[64,65] or Ishan Amin and co-workers' development of student–teacher model-distillation strategies.[97] In the future, cross-pollination between these different efforts will hopefully lead to the development of models which inherit the best characteristics of both.

Egret-1 is limited to predictions of energies, forces, and derivatives thereof (like frequencies). While these are by far the most commonly simulated properties, the lack of dipole moments, atom-centered charges, and other electronic properties limits the applicability of these models to certain tasks. Egret-1 also is only trained on gas-phase calculations and thus predicts properties only in the gas phase; we plan to address this limitation in the future, either through fine-tuning or by using an auxiliary solvent-correction model like GNNIS.[116,117]

Our work also illustrates important outstanding challenges in the NNP field. Unlike many domains of machine learning, atomistic simulation appears to not yet have reached a scale in which increasing dataset size has a uniformly positive effect. In many cases, adding data could be shown to reduce the performance of the model on virtually every test set, including benchmarks which appear similar to the new data. Even in cases when adding more data improved relevant benchmark scores, these improvements were often offset by losses in some other area—for instance, Egret-1e is best at thermochemistry, but produces significantly inferior frequencies. This "zero-sum" model behavior may suggest that more expressive architectures are needed for further systematic improvements—or simply that the scale of the Egret-1 models is insufficient.

We hypothesize that better training strategies and dataset-aggregation strategies will be key to continued improvement with today's datasets. While all of the models described in this work use a single-phase training procedure, more complex pretraining and fine-tuning protocols may make it possible to increase diversity by introducing new data distributions without causing catastrophic losses in general performance. We note that our observations are limited to a single architecture, and that dataset sensitivity may differ across architectures.

This study highlights the need for better ways to quantify diversity and similarity in training datasets. Here, we were able to identify productive combinations of existing datasets through essentially a trial-and-error approach, which quickly becomes expensive, time-consuming, and impractical. Data-driven methods to guide the construction of new, well-behaved datasets would be immensely valuable for future efforts, particularly because adding even small amounts of new data to a well-behaved dataset can result in substantial performance degradation. For example, simple heuristics about the optimal ratio of near-equilibrium to far-from-equilibrium structures to include, the optimal number of conformers per molecule, the best way to include transition states and other reaction-path structures, and so on would vastly simplify the process of creating new datasets and training more general NNPs moving forward.[118]

Our experience training the Egret-1 models also reinforces the importance of evaluating non-energy-based benchmarks throughout the training process. Based solely on energy benchmarks, we might easily have concluded that Egret-1e or a similar model was best, and discovered only later on that the gradients and frequencies were inaccurate—the sensitivity of these higher-order properties of the potential-energy surface to dataset construction makes it imperative to continuously monitor them. We observed a similar phenomenon when training EquiformerV2 models on this dataset.[119]

# 5 Conclusion

In this work, we report the Egret-1 family of neural network potentials, which aim to achieve zero-shot chemical accuracy for closed-shell bioorganic simulation tasks. Our benchmarks show that Egret-1 achieves this goal for many important simulation tasks—and, in many circumstances, Egret-1 is more accurate than the small-basis-set quantum-chemical methods routinely employed in academic research, drug discovery, and materials science. The models presented here can thus serve as replacements for quantum chemistry in many contexts, even without system-specific fine-tuning.

The Egret-1 models are available at github.com/rowansci/egret-public under an MIT license and can also be run through the Rowan computational-chemistry platform. While we have spent considerable time benchmarking these models on a wide variety of challenging tasks, we expect that broader scientific usage will be indispensable in driving further improvements, and look forward to learning more about these models' strengths and weaknesses.

The Egret-1 models have several obvious lacunae—such as being limited to relatively few elements and neutral closed-shell molecules in the gas phase—which we plan to address in the future. These models are also too slow for many important workflows, like molecular dynamics studies of large biomolecules or polymers, which limits their applicability and utility. Thus, while we feel this work represents an important step towards the "holy grail" of fast, accurate, and reliable molecular simulation,[2] many more steps will be needed to fully achieve the goals outlined in the introduction. In particular, we anticipate that a combination of improved dataset scale and quality, more expressive architectures, and performance optimization will make it possible to achieve significantly improved accuracy, speed, and generality, which we expect to have a substantial impact on discovery across the chemical sciences.

## Acknowledgements

The authors thank Keir Adams, Justin Airas, Ilyes Batatia, Simon Batzner, Cristian Bodnar, Jackson Burns, Tim Duignan, Zach Fried, Joe Gair, Ishaan Ganti, Kevin Greenman, Chandler Greenwell, Michael Hla, Patrick Hsu, Bowen Jing, David Klee, Veljko Kovac, Tony Kulesa, Eugene Kwan, Eli Laird, Yi-Lun Liao, Abhishaike Mahajan, Narbe Mardirossian, Alex Mathiasen, Tom McGrath, Gabriel Mongaras, Albert Musaelian, Mark Neumann, Vedant Nilabh, Sam Norwood, John Parkhill, Andrew Rosen, Marcus Sak, Justin Smith, Guillem Simeon, Hannes Stärk, Kayvon Tabrizi, Zach Ulissi, Nick Wall & Larry Zitnik for helpful discussions, and Tim Duignan, Joe Gair, Michael Hla, Alex Mathiasen, Mark Neumann, & Marcus Sak for editing early drafts of this manuscript.

# Appendix: Hyperparameters

| **Hyperparameter** | **Egret-1** | **Egret-1e** | **Egret-1t** |
|---|---|---|---|
| max_epochs | 190 | 190 | 190 |
| batch_size | 128 | 128 | 128 |
| val_split | 0.2 | 0.2 | 0.2 |
| r_max | 6 | 6 | 6 |
| atomic_inter_scale | 1 | 1 | 1 |
| atomic_inter_shift | 0 | 0 | 0 |
| num_bessel | 8 | 8 | 8 |
| num_polynomial_cutoff | 5 | 5 | 5 |
| max_ell | 3 | 3 | 3 |
| num_interactions | 2 | 2 | 2 |
| num_elements | 10 | 11 | 10 |
| num_channels | 192 | 192 | 192 |
| max_L | 2 | 2 | 2 |
| correlation | 3 | 3 | 3 |
| gate | silu | silu | silu |
| pair_repulsion | False | False | False |
| radial_type | Bessel | Bessel | Bessel |
| distance_transform | None | None | None |
| learning_rate | 0.01 | 0.01 | 0.01 |
| weight_decay | 5e-10 | 5e-10 | 5e-10 |
| gradient_clip_val | 1 | 1 | 1 |
| lr_scheduler_gamma | 0.99 | 0.99 | 0.99 |
| lr_scheduler_patience | 20 | 20 | 20 |
| w_energy | 40 | 40 | 40 |
| w_forces | 1000 | 1000 | 1000 |
| swa | True | True | True |
| swa_lr | 0.00025 | 0.00025 | 0.00025 |
| start_swa | 115 | 115 | 115 |
| swa_w_energy | 1000 | 1000 | 1000 |
| swa_w_forces | 10 | 10 | 10 |
| ema | True | True | True |
| ema_decay | 0.9995 | 0.9995 | 0.9995 |
| avg_num_neighbors | 22.912 | 22.642 | 20.033 |

Table 15: Hyperparameters for Egret-1 models

| Element | Atomic Energy (eV) |
|---|---|
| H | −13.571 965 |
| C | −1030.567 165 |
| N | −1486.375 026 |
| O | −2043.933 693 |
| F | −2715.318 529 |
| Si[a] | −7876.342 032 |
| P | −9287.407 133 |
| S | −10 834.484 471 |
| Cl | −12 522.649 269 |
| Br | −70 045.283 851 |
| I | −8105.734 201 |

Table 16: Atomic energy hyperparameters for Egret-1 models
[a]Only for Egret-1e